\documentclass[]{spie}
 
\usepackage{amsmath,amsfonts,amssymb}
\usepackage{graphicx}
\usepackage[colorlinks=true, allcolors=blue]{hyperref}
\usepackage{booktabs}
\usepackage{tikz}

\title{The Optical Design Concept for the Atacama Large Aperture Submillimeter Telescope (AtLAST)}

\author[a]{Patricio A. Gallardo}
\author[b]{Roberto Puddu}
\author[c]{Tony Mroczkowski}
\author[d]{Martin Timpe}
\author[d]{Pierre Dubois-dit-Bonclaude}
\author[d]{Manuel Groh}
\author[d]{Matthias Reichert}
\author[e]{Claudia Cicone}
\author[f]{Hans J. Kaercher}

\affil[a]{Kavli Institute for Cosmological Physics, The University of Chicago, USA}
\affil[b]{Instituto de Astrofísica and Centro de Astro-Ingeniería, Facultad de Física, Pontificia Universidad Católica de Chile, Santiago, Chile}
\affil[c]{European Southern Observatory, Karl-Schwarzschild-Str.\ 2, Garching 85748, Germany}
\affil[d]{OHB Digital Connect, Weberstra\ss e 21, D-55130 Mainz, Germany}
\affil[e]{Institute of Theoretical Astrophysics, University of Oslo, PO Box 1029, Blindern 0315, Oslo,
Norway}
\affil[f]{Independent Consultant, Kirchgasse 4, D-61184 Karben, Germany}

\begin{document}

\maketitle

\begin{abstract}
    The Atacama Large Aperture Submillimeter Telescope (AtLAST) aims to be the premier next generation large diameter (50-meter) single dish observatory capable of observations across the millimeter/sub-millimeter spectrum, from 30 to 950~GHz. The large primary mirror diameter, the 2-degree field of view and its large 4.7-meter focal surface give AtLAST a high throughput (aperture size times field of view) and grasp (throughput times spectral reach), with the ability to illuminate $>\mathcal{O}(10^7)$ detectors. The optical design concept for AtLAST consists of a numerically optimized two-mirror Ritchey-Chrétien system with an additional flat folding mirror, which enables a quick selection among its planned six instrument positions. We present the optical design concept and discuss the expected optical performance of AtLAST. We then present design concepts that can be implemented in the receiver and instrument optics to correct for astigmatism and mitigate the high degree of curvature of the focal surface in order to recover significant fractions of the geometric field of view at sub-millimeter wavelengths.
\end{abstract}


\section{Introduction}\label{sec:introduction}

The Atacama Large Aperture Submillimeter Telescope, AtLAST, is a proposed fifty-meter telescope, capable of illuminating a 4.7-meter focal surface with a two-degree field of view. AtLAST aims to be the premier wide-field mapping telescope at frequencies spanning 30-950 GHz (1 cm - 350 $\mu$m), capable of diffraction-limited observations with resulting resolutions of a few to a few tens of arcseconds across the millimeter/sub-millimeter bands. The large aperture, large field of view and large spectral grasp will give AtLAST the ability to observe the microwave sky with unparalleled mapping speeds and resolution.

The scientific objectives of AtLAST are broad and range from resolved observations of the Sun, solar system objects, molecular clouds and star forming regions, nearby galaxies, and large scale cosmic structures, as well as detections of protoplanetary discs in our Galaxy and distant dusty star forming galaxies out to their redshift of formation.\cite{Ramasawmy2022, Cordiner2024atlast, DiMascolo2024, Klaassen2024atlast, Lee2024atlast, Liu2024atlast, orlowskischerer2024atlast, vanKampen2024atlast, Wedemeyer2024atlast}\footnote{The AtLAST science cases can be found in the Open Research Europe collection, located here: \url{https://open-research-europe.ec.europa.eu/collections/atlast}.} A broader description of the AtLAST design has been presented separately in Mroczkowski et al.\ 2024 \cite{Mroczkowski2024}, while the scientific motivations are presented in Booth et al.\ 2024 (submitted to these proceedings)\cite{Booth2024}. These proceedings also include additional subsystem descriptions for the telescope structure (Reichert et al.\ submitted)\cite{Reichert2024} and an energy recovery system (Kiselev et al.\ submitted)\cite{Kiselev2024}, while Puddu et al.\ (submitted) presents results of full-wave physical optics simulations for the optical design discussed here.

This document describes the optical design parameters of the fifty-meter AtLAST, including aperture definitions and image quality metrics. A basic instrument consisting of three silicon lenses with biconic lens corrections is presented to demonstrate a suitable instrument concept that can serve as a starting point of future receiver development. Section \ref{sec:opticaldesign} presents the telescope optical design parameters. Section \ref{sec:opticalperformance} presents the optical evaluation of this design. In Section \ref{sec:cameraconcept} we present a suitable concept for a three lens camera instrument with biconic lenses. Finally, we conclude in Section \ref{sec:conclusions}.

\begin{figure}[tbh]
    \centering
    \includegraphics[height=0.38\textwidth, trim={0cm 0cm 0cm 1cm},clip]{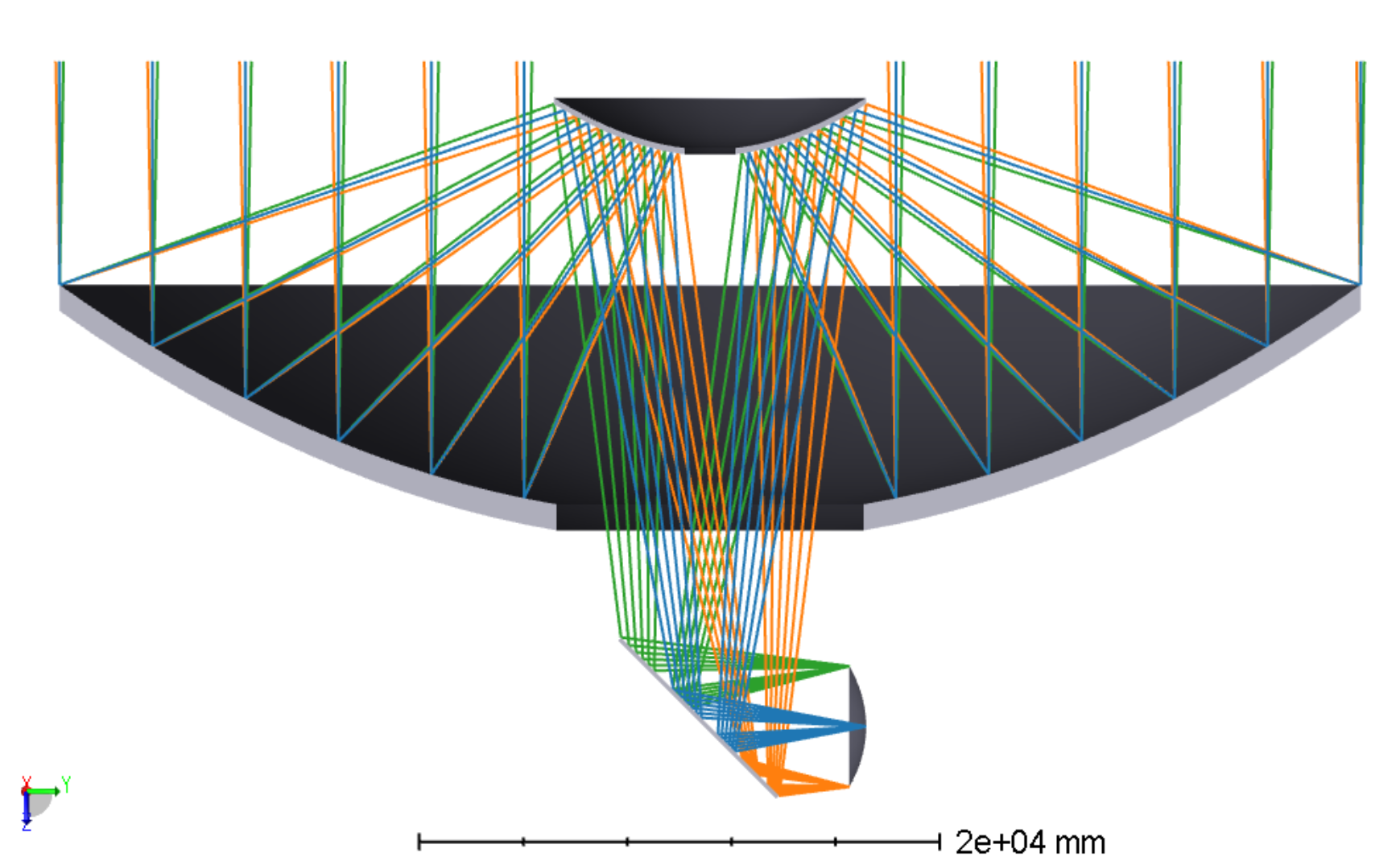}
    \includegraphics[height=0.38\textwidth, trim={28cm 2cm 42cm 3cm},clip]{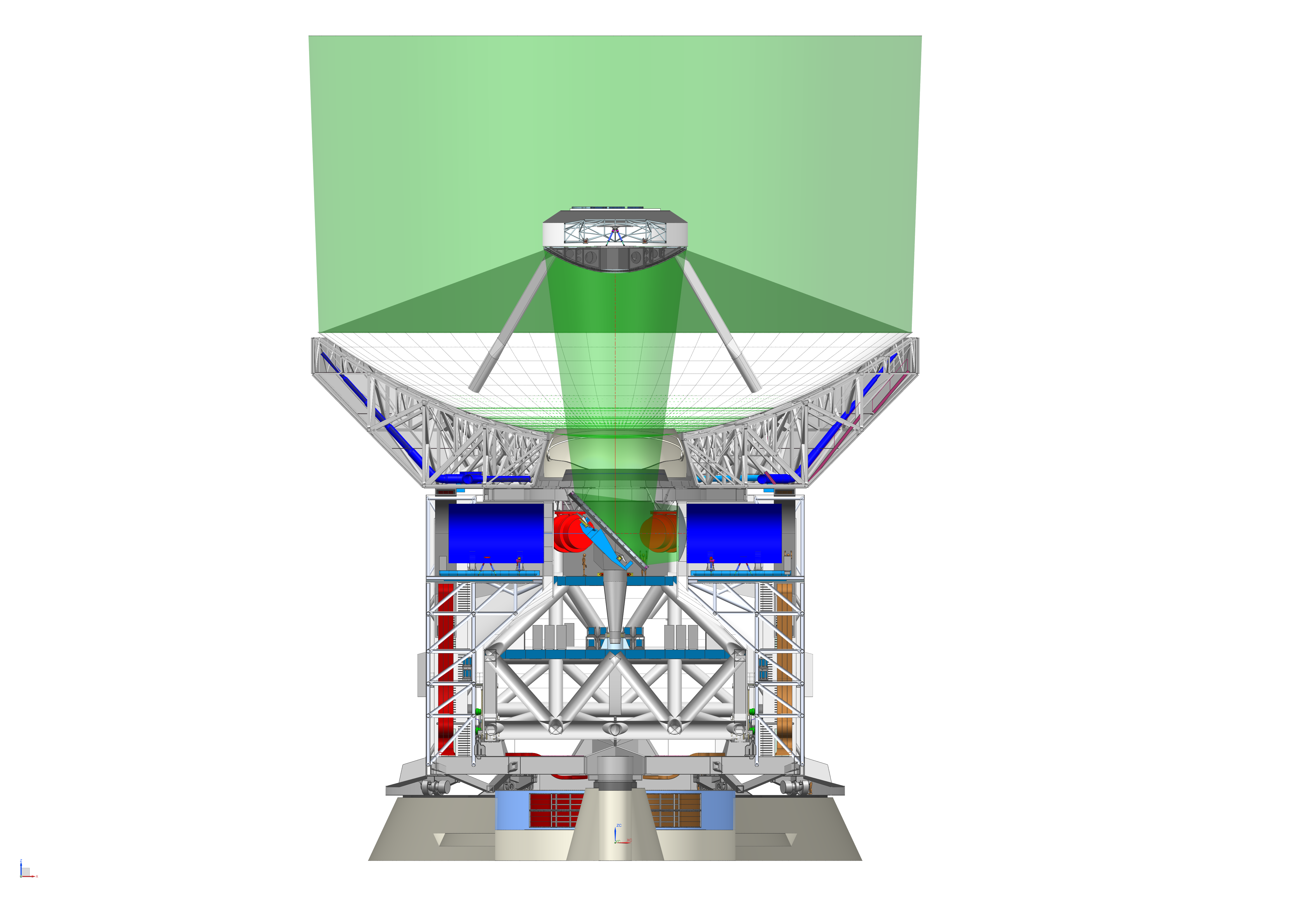}
    \caption{\textbf{Left:} Optical layout of the Atacama Large Aperture Submillimeter Telescope, AtLAST. Rays shown from the sky (top) reflect on the 50-meter primary mirror (M1), reflect once more on a 12-meter secondary mirror (M2) to reflect on an 8.7 tertiary mirror (M3), to finally focus on a 4.7 meter focal surface (FS). Fields shown make an angle of $\pm 1$ degree from the boresight (2-degree full field of view). Model includes the secondary aperture obstruction. \textbf{Right:} Cross-section of AtLAST pointing towards the zenith to replicate the orientation shown in the ray trace (left). }
    \label{fig:optlayout}
\end{figure}

\section{Optical Design Parameters}\label{sec:opticaldesign}

The telescope is composed of two curved surfaces in a Ritchey-Chrétien configuration and a flat folding mirror that enables selection among the multiple planned instrument locations \cite{AtLAST_memo_1}. The 50-meter diameter primary (M1) illuminates a 12-meter diameter secondary (M2), which in turn illuminates a flat $6.2 \times 8.8\,$~meter fold mirror (M3). Light from the fold mirror focuses into a curved focal surface (FS). The optical layout is shown in Figure~\ref{fig:optlayout}. Mirror shapes are given by a standard conic surface of revolution $z(r)$ according to 
\begin{equation}
    \label{eq:mirrorshapes}
    z(r) = \frac{r^2/R}{1 + \sqrt{1-(1+k)(r/R)^2}},
\end{equation}
where $\rm R$ is the radius of curvature of the surface, and $\rm k$ is the conic constant. The z-axis for the primary points in the direction from the sky to the primary, while for the secondary it points from the secondary to the tertiary and for the focal surface it points from the tertiary to the focal surface. Numerical parameters for the evaluation of Equation \ref{eq:mirrorshapes}, mirror positions and apertures are given in Table \ref{tab:mirrorprescription}.

\begin{table}[!ht]
    \centering
    \centering
\begin{tabular}{lccccccccc}
\toprule
{} &    R &     k &  x &     y  &   z &  $R_{\rm min}$&  $R_{\rm max}$  &$R_x$ &$R_y$ \\
\midrule
Primary (M1)   & -35000.00 & -1.01016   &    0.00 &    0.00 & 17500.00 &       5900 &      25000 & -- & --\\
Secondary (M2) &  -8331.68 & -1.76177   &    0.00 &    0.00 &  3624.25 &        980 &       6000 & -- & --\\
Tertiary (M3)  &       $\infty$ &  0.00 &    0.00 & -383.35 & 25116.65 &   --    &    --   & 3050 & 4350\\
Focal Surface (FS)     &   4243.85 & -0.79798   &    0.00 & 6000.00 & 25500.00 &          -- &       2350 & -- & --\\
\bottomrule
\end{tabular}
    \caption{Mirror shape parameters (R and k), locations (x, y, z), circular active optical apertures and obscurations ($R_{\rm min}$, $R_{\rm max}$) and elliptical M3 aperture $R_x$ $R_y$ for the AtLAST design. The radius of curvature is denoted by $\rm R$, the conic constant is given by $k$. Positions of surfaces in 3D space are given by x, y, z. Tertiary surface is defined by an elliptical aperture, with semi-diameters given by $R_x$ and $R_y$. Units for all columns are mm, except for the conic constant, which is unit-less.}
    \label{tab:mirrorprescription}
\end{table}

\section{Telescope Optical Performance}\label{sec:opticalperformance}
In this section we detail figures of merit that describe the expected performance of the telescope optical design such as diffraction-limited field of view, spot diagrams, the Huygens point spread function, f-numbers and mirror footprints.
 
\begin{figure}[ht]
    \centering
    \begin{tikzpicture}
    \draw (0, 0) node[inner sep=0] {\includegraphics[width=0.55\textwidth]{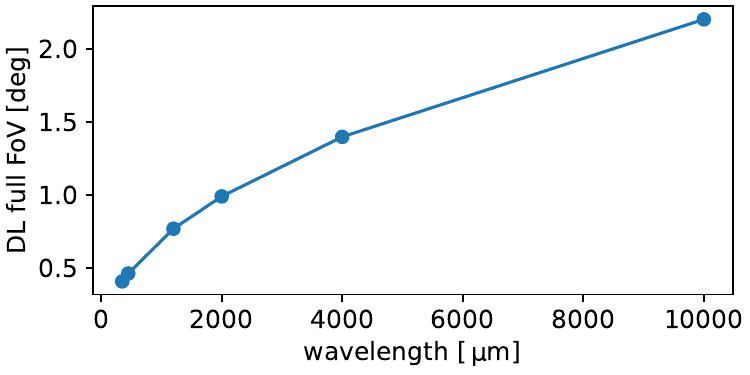}};
    \draw (-4.3, 2.2) node {a)};
    \end{tikzpicture}
    \begin{tikzpicture}
    \draw (0, 0) node[inner sep=0] {
    \includegraphics[width=0.42\textwidth, trim={2cm 8cm 10cm 1cm}, clip]{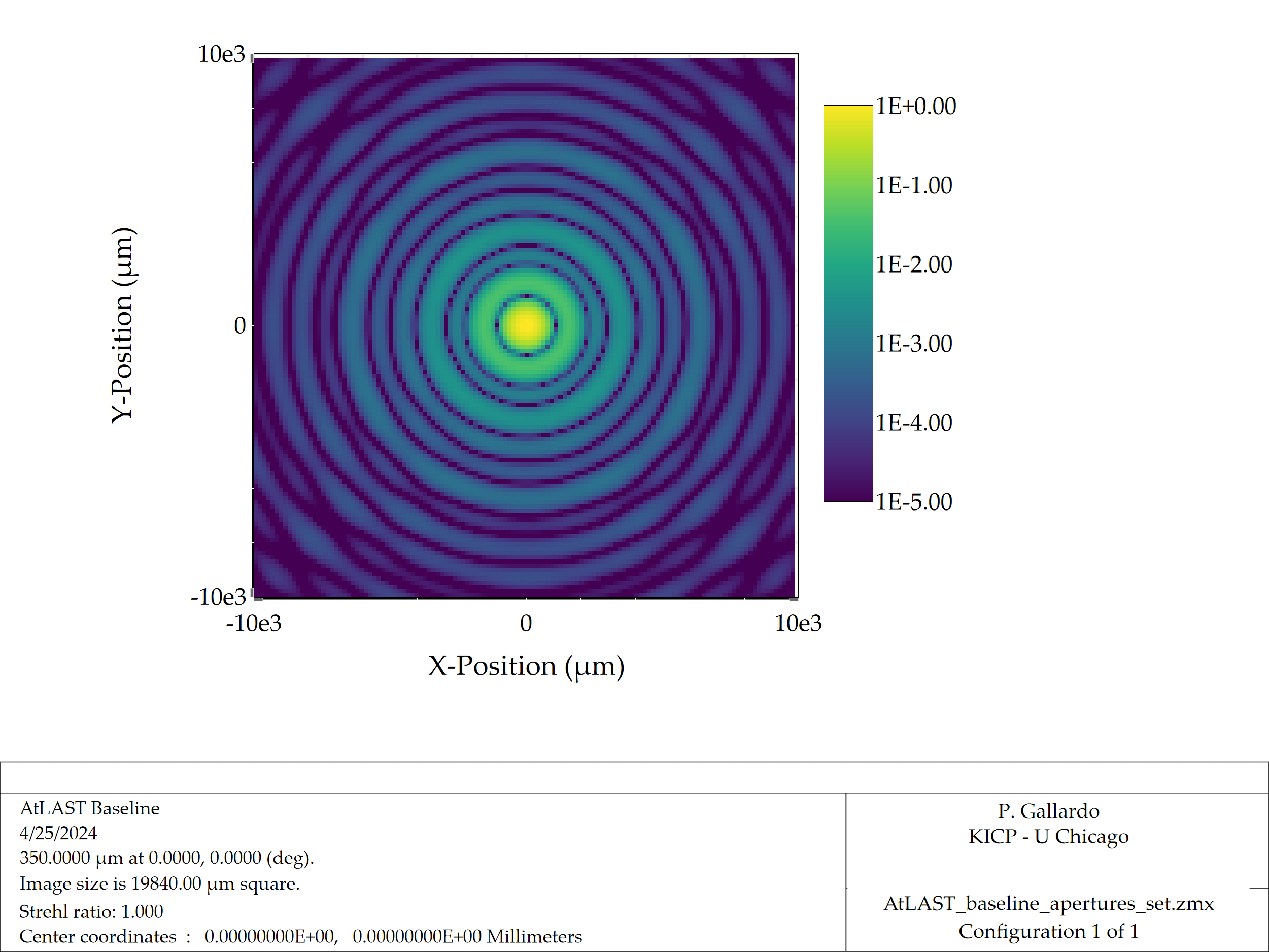}};
    \draw (-3.4, 2.2) node {b)};
    \end{tikzpicture}

    \begin{tikzpicture}
    \draw (0, 0) node[inner sep=0] {
    \includegraphics[width=0.33\textwidth, trim={8cm 8cm 10cm 2cm}, clip]{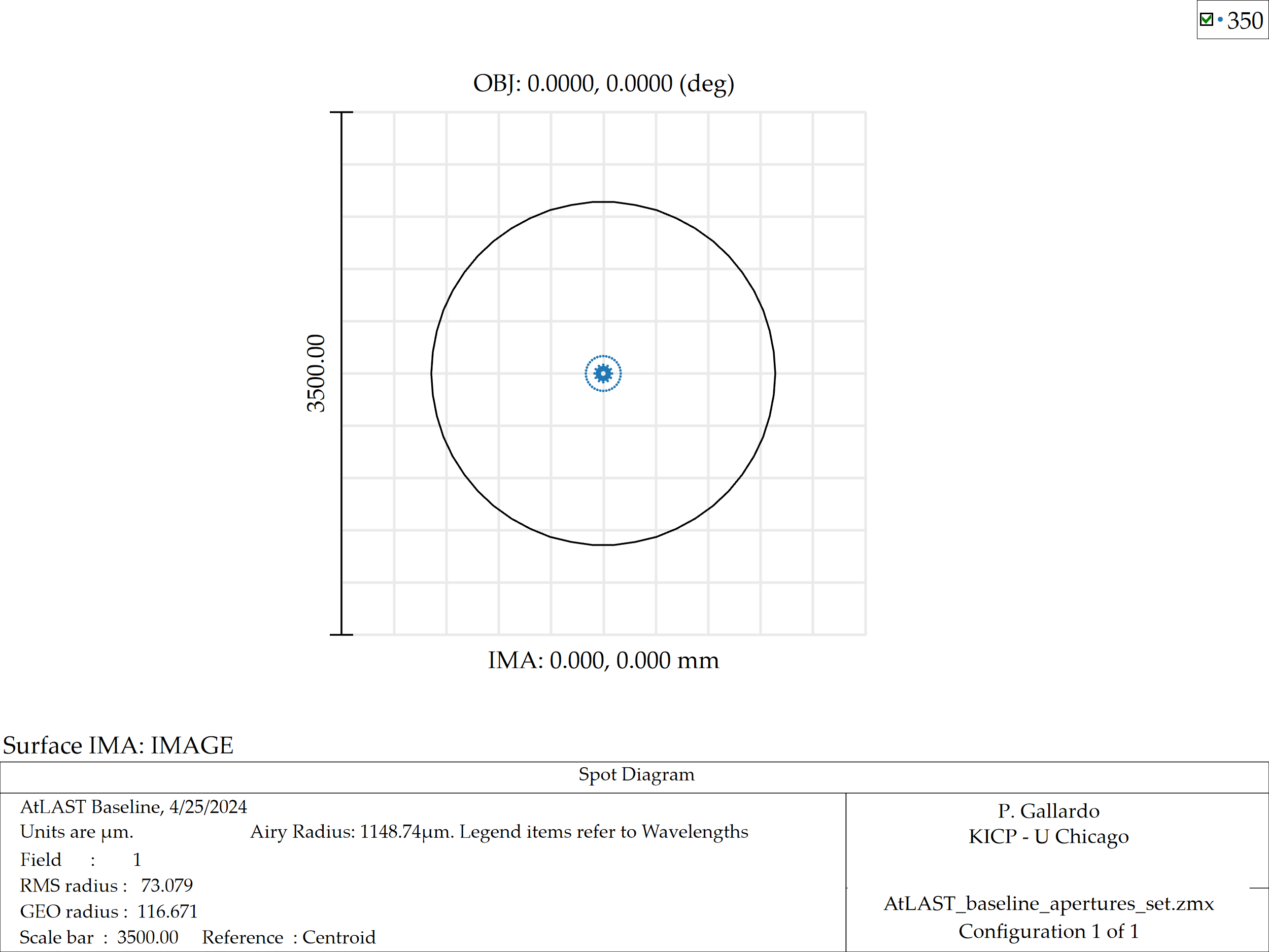} \includegraphics[width=0.33\textwidth, trim={8cm 8cm 10cm 2cm}, clip]{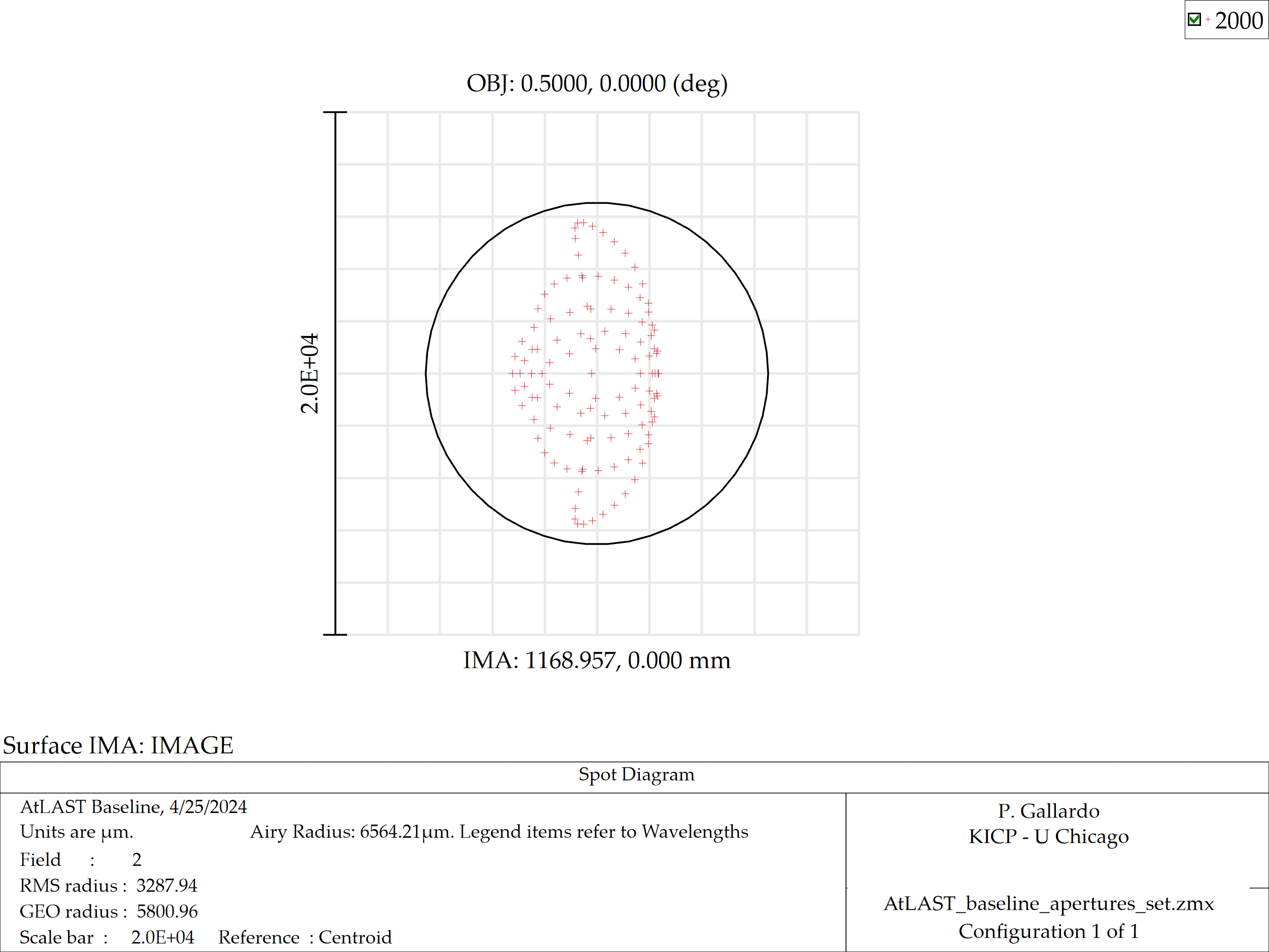} \includegraphics[width=0.33\textwidth, trim={8cm 8cm 10cm 2cm}, clip]{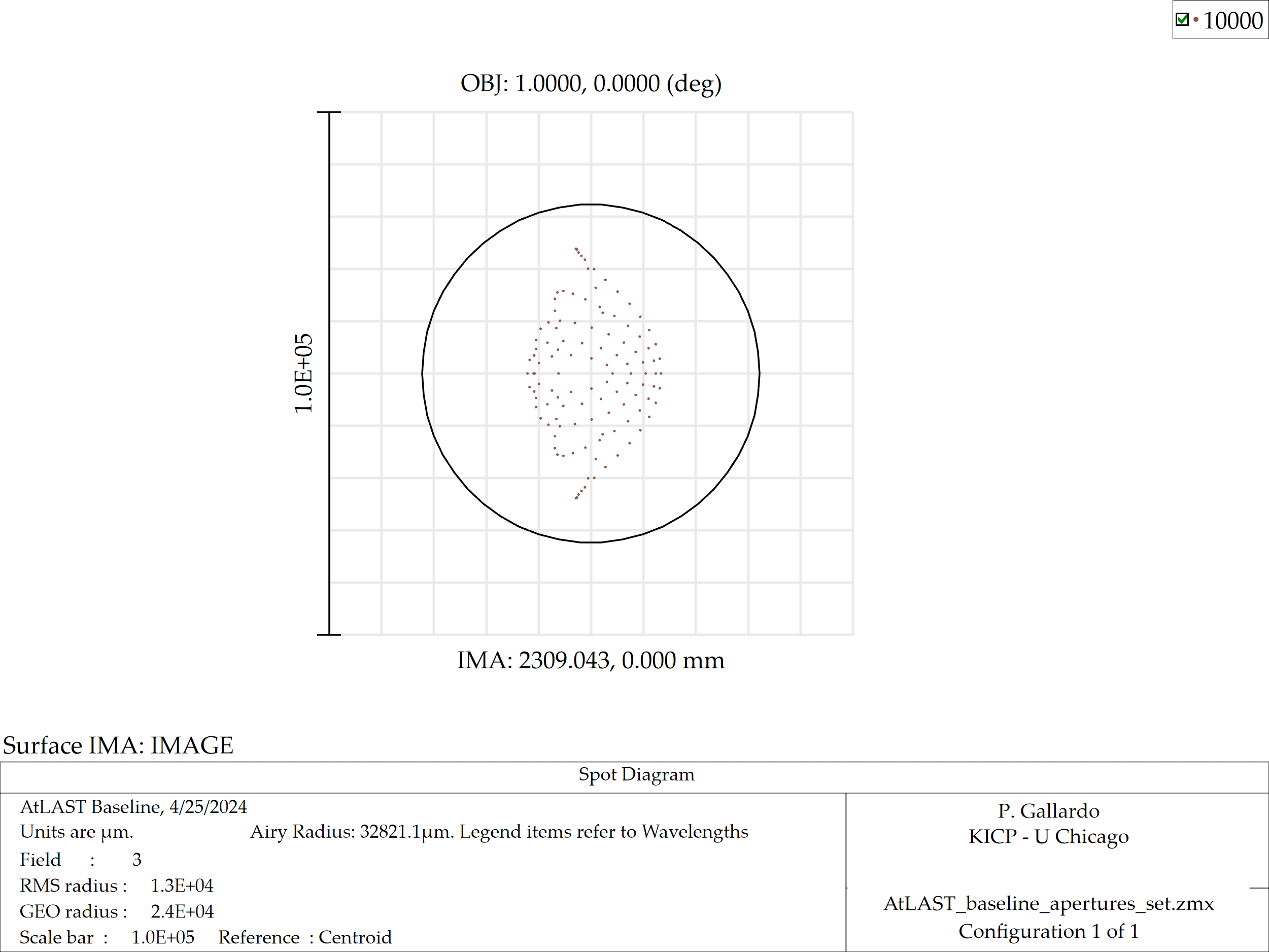}};
    \draw (-8.7, 2.2) node {c)};
    \end{tikzpicture}

    \caption{Image quality metrics for the telescope without re-imaging optics. a) Diffraction-limited full field of view of the telescope as a function of wavelength. The diffraction-limited performance at $350 \,\rm \mu m$ covers 0.4 degrees on the sky, which corresponds to 0.9 meters in diameter at the telescope focal surface. b) Simulated beam at the center field at a wavelength of $\lambda = 350\, \rm \mu m$. The double ring structure is produced by the obstruction due to the secondary mirror, which produces oscillations that are superposed to the ringing of the unobstructed optical system. This model includes the effect of the obstruction of four spiderweb legs with a thickness of $245\, \rm mm$, which cover the primary mirror up to a radius of 16.7 meters. c) Spot diagrams at wavelengths of $350\,\mu\rm m$, 2.0\,mm and 10\,mm for the fields located at 0, 0.5 and 1.0 degrees from the boresight with corresponding Strehl ratios of 0.99, 0.78 and 0.86. The black circle represents the diffraction limit at these wavelengths, corresponding to the first null in the Airy disk (${1.22\lambda}/{D}$).  We note that the linear dimension has been scaled with wavelength and is shown in the figure in microns.}
    \label{fig:dlfov}
\end{figure}

The full diffraction-limited field of view is quantified computing Strehl ratios from $350\, \rm \mu m$ to $10\, \rm mm$. The diffraction-limited field of view for the telescope is the maximum angle at which the Strehl ratio stays above 0.8. Figure~\ref{fig:dlfov} (a) shows the full diffraction-limited field of view of the telescope without re-imaging cameras as a function of wavelength. This telescope-only field of view can be interpreted as a lower bound for the achievable field of view as the reimaging optics can correct some aberrations as discussed in Section \ref{sec:cameraconcept}. Figure~\ref{fig:dlfov} (b) shows a Huygens point spread function computed using Ansys Zemax OpticStudio (\footnote{See
\href{https://www.ansys.com/products/optics/ansys-zemax-opticstudio}{https://www.ansys.com/products/optics/ansys-zemax-opticstudio}.}) at $\lambda = 350 \, \mu\rm m$. The concentric rings are an indication that the system reaches the diffraction limit at this wavelength. The double ring structure, with alternating thick and thin rings, is produced by the obstruction of the aperture created by the secondary mirror, and the non-uniform width of the rings is produced by the fourfold symmetry of the obstructing leg struts that support the secondary mirror.  An accompanying proceedings paper by Puddu et al. in this series shows more detailed physical optics beam calculations. Figure~\ref{fig:dlfov} (c) shows the spot diagrams for three field positions (0, 0.5, and 1.0 degrees from the boresight) and the Airy disk size for three wavelengths (0.35, 2.0, and 10~mm).

\begin{figure}
    \centering
    \includegraphics[width=0.48\textwidth]{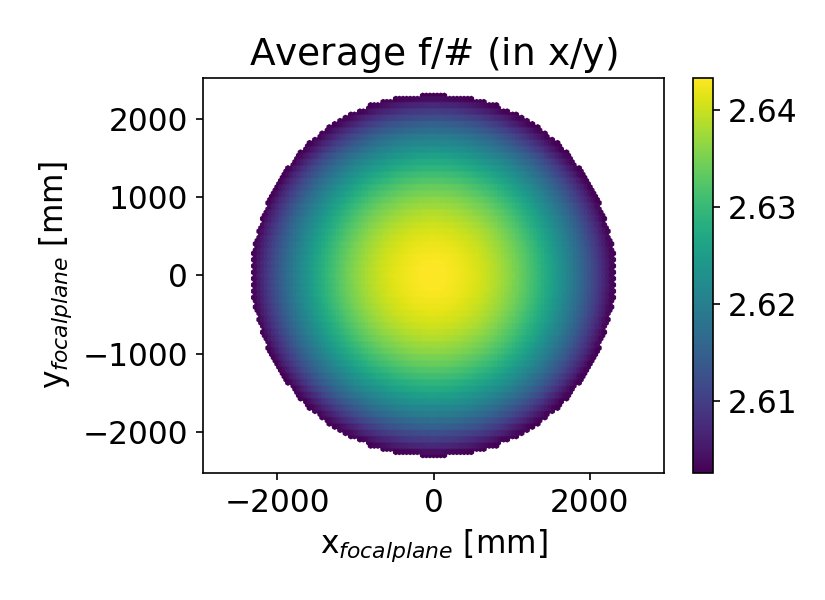}
    \includegraphics[width=0.48\textwidth]{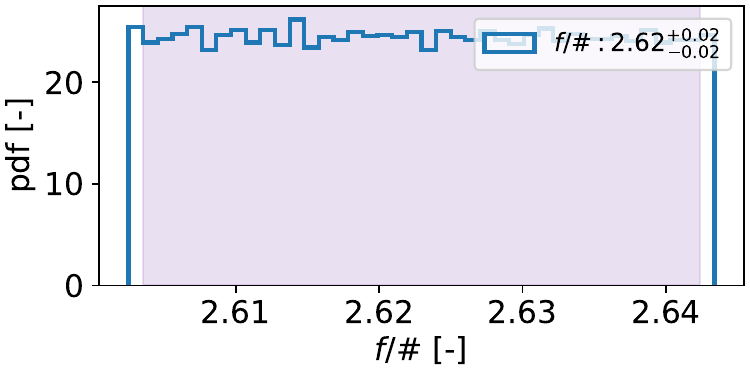}
    \caption{Left: Distribution of f-numbers in the telescope focal plane surface. Right: Distribution of f-number with its 95\% interval values.}
    \label{fig:f_numbers}
\end{figure}

The f-number for the telescope system is quantified by the angle of arrival of the marginal ray to the focal surface relative to the chief ray. Using this convention, the f-number is $2.62 \pm 0.02$ in $95\%$ of the focal plane. Figure~\ref{fig:f_numbers} shows the f-number distribution across the focal surface (left) and a histogram over the focal surface area (right). This focal number is chosen to enable compatibility with planned and under construction instruments, like the instruments for the Simons Observatory, CCAT and CMB-S4. \cite{Parshley2018, Chapman2022, Vavagiakis2022, Gallardo2024}

\begin{figure}
    \centering
    \includegraphics[width=0.24\textwidth, trim={9cm 8cm 11cm 3cm},clip]{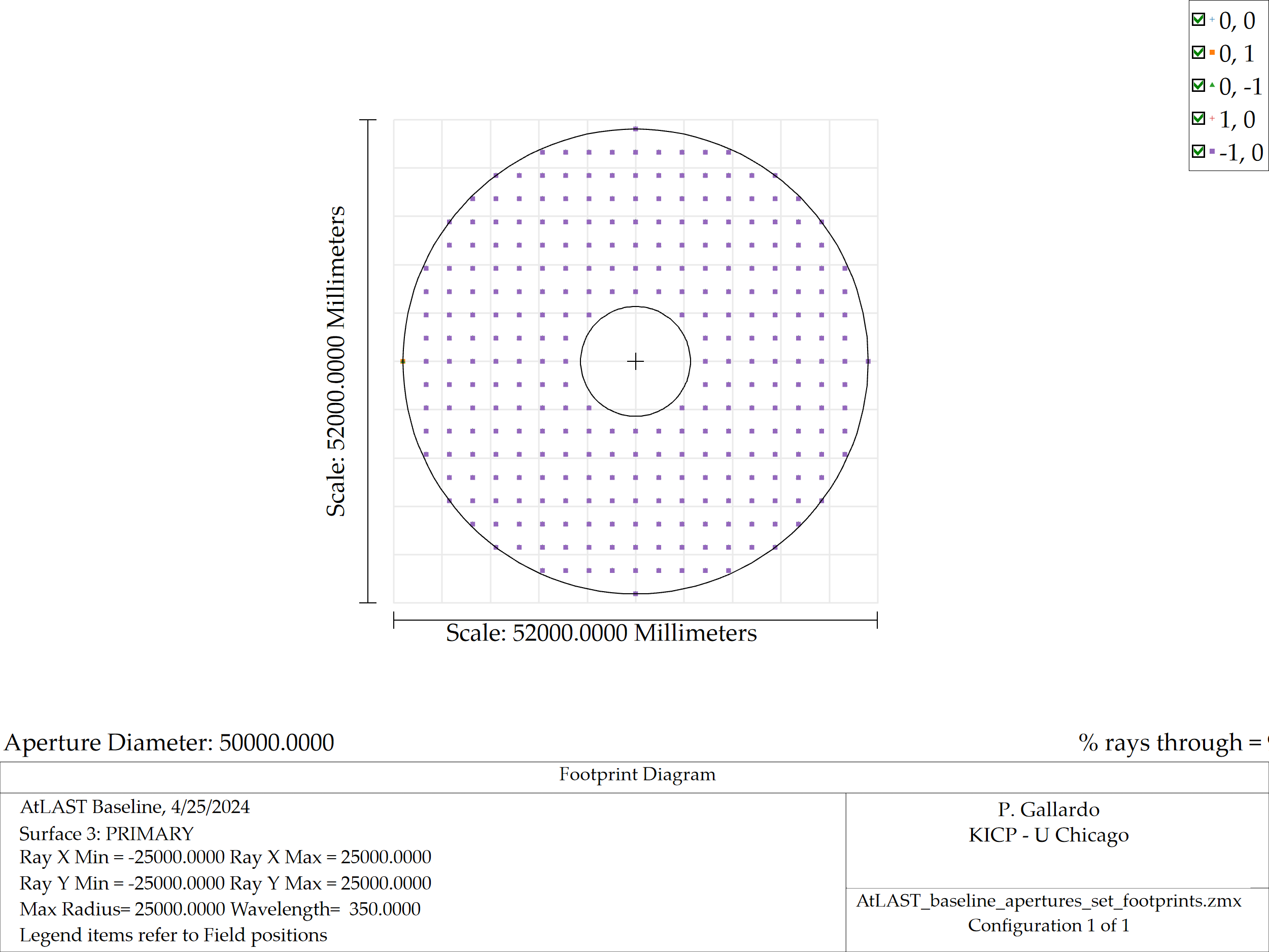}
    \includegraphics[width=0.24\textwidth, trim={9cm 8cm 11cm 3cm},clip]{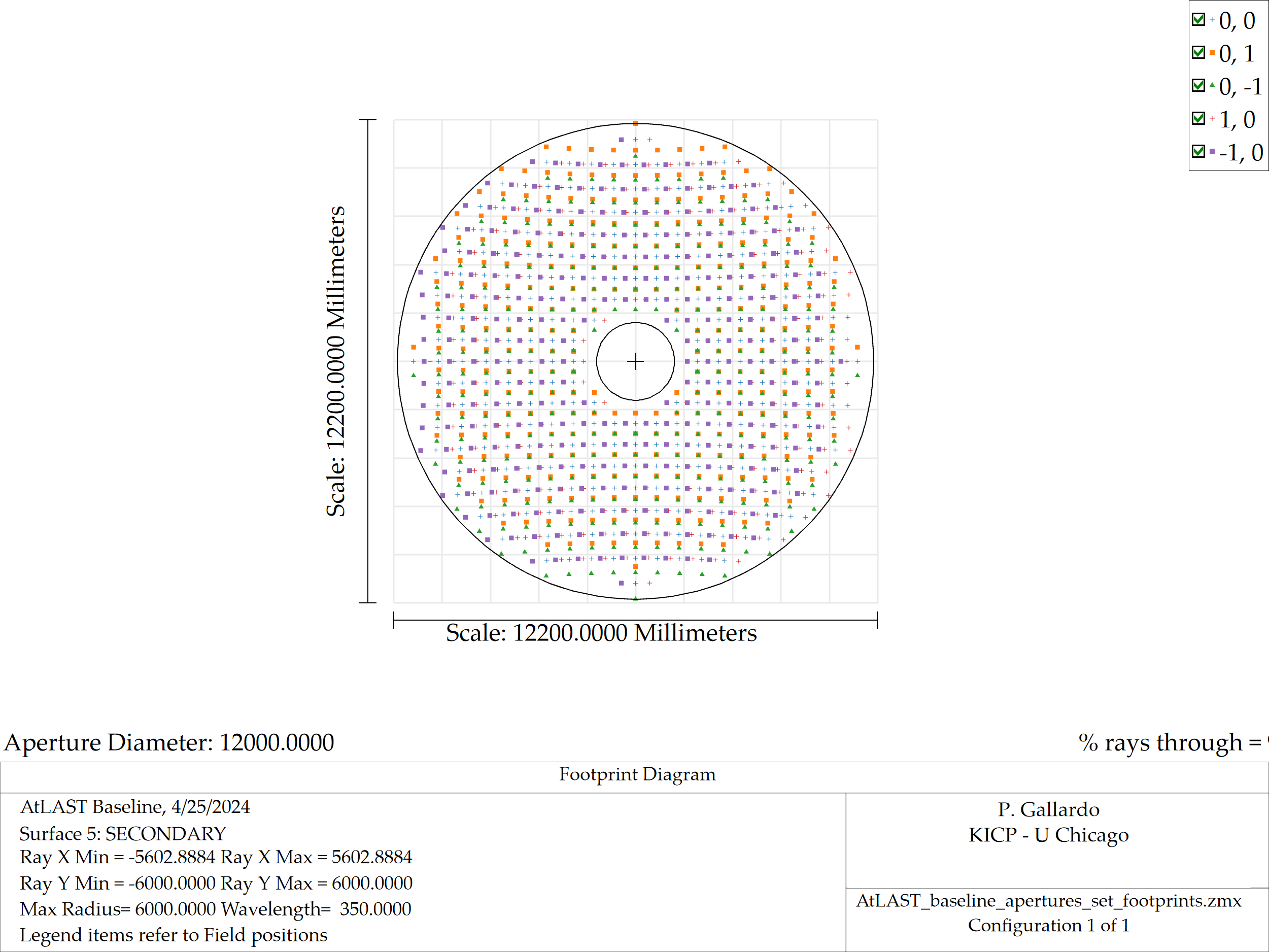}
    \includegraphics[width=0.24\textwidth, trim={9cm 8cm 11cm 3cm},clip]{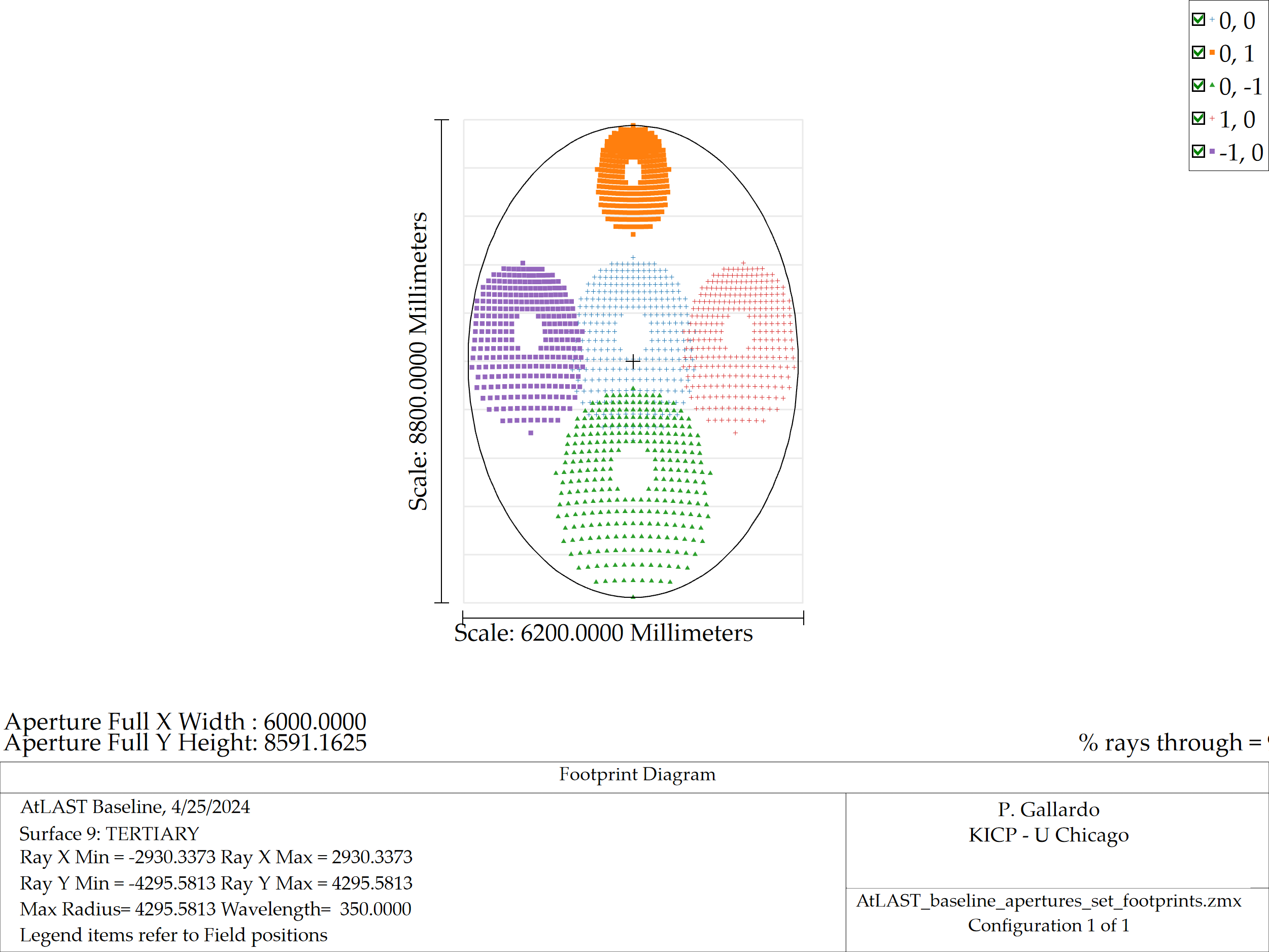}
    \includegraphics[width=0.24\textwidth, trim={9cm 8cm 11cm 3cm},clip]{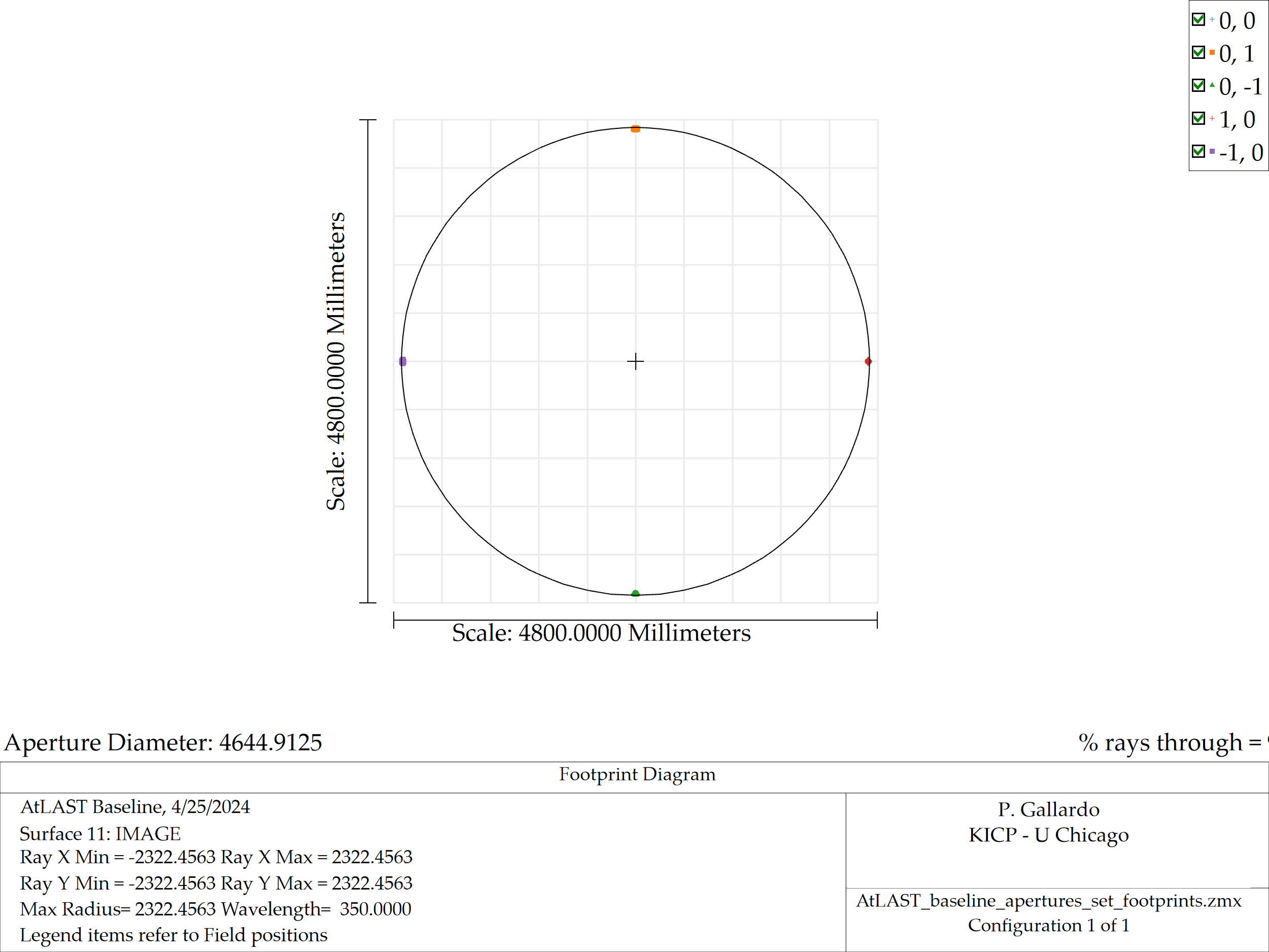}
    
    \caption{Optically active footprints on telescope surfaces. From left to right: i) Primary mirror and the central obstruction produced by the twelve-meter secondary. ii) Secondary mirror footprint, showing the central non-illuminated region  produced by the projection of the secondary obstruction on the primary mirror for a 2-degree field of view. iii) Tertiary flat mirror. iv) Focal plane. Surface apertures are given by Table \ref{tab:mirrorprescription}. }
    \label{fig:footprints}
\end{figure}

Footprints for the three mirrors and focal surface are shown in Figure~\ref{fig:footprints}. The primary mirror obstruction (given by the 12-meter diameter secondary) shown in the leftmost panel defines the size of the active area in the secondary (second panel). The circular converging cone of light coming from the secondary is intersected by the tilted flat tertiary mirror, which causes the  illumination of the tertiary mirror to be elliptical, with bigger ellipses on the section of the mirror that is the closest to the secondary mirror, while the smallest ellipses are in the region of the mirror that is the furthest from the secondary, as shown in the third panel. The rightmost panel of Figure~\ref{fig:footprints} shows the focal surface footprints for fields at 1 degree from the boresight, note that a 2-degree field of view illuminates 4.7~meters at the focal surface.

\section{Camera Concept}\label{sec:cameraconcept}

Although the development of instruments for AtLAST will be studied in the future in response to specific scientific cases and their requirements, this section describes a basic instrument concept composed of an array of cameras, which can be evolved further in order to populate the focal plane with instruments. This concept is intended as a starting point for future development, and will most certainly be optimized further in the future.

The large focal plane of AtLAST needs to be divided into smaller optics tubes of tens of centimeters in diameter according to current technologies. These optics tubes contain reimaging optics, filters and cryogenic detectors. To demonstrate the ability of populating the focal surface with optics tubes like these, we use a minimal three-lens camera. The three lenses in the system can be made out of silicon, this concept is common in ground-based cosmic microwave background experiments \cite{Dicker2018,Gallardo2024}. The first two lenses form an image of the primary mirror on an intermediate cold stop (also called a `Lyot' stop), this optical configuration allows to set the aperture of the primary mirror and block unwanted thermal radiation from room temperature surroundings.

The camera concept chosen in this demonstration uses an array of $40$ and $20\,\rm {cm}$ diameter cameras arranged in a hexagonal lattice pattern. The 40 $\, \rm cm$ camera size corresponds to one of the largest optical quality silicon formats that is easily commercially available at present time, and is chosen to maximize the area coverage with a minimum number of 85 tubes. We note that smaller camera concepts can be beneficial to optimize the system for image quality, as a smaller field lens allows less variation in wavefront error, which allows the optimization to find better solutions. For this reason, it is foreseeable that the highest frequency cameras of AtLAST could be made of smaller optics (of perhaps 20 or 30 cm diameter) to increase the image quality at the detector plane. Another possible extension of the three-camera concept presented here, is the use of a fourth lens, which gives more degrees of freedom to correct optical aberrations.

The curvature of the focal surface in AtLAST constrains the orientation of the camera windows. In the concept presented here the orientation of the camera was chosen to be perpendicular to the focal surface at the center of the window. For simplicity, a coaxial design was chosen, where all lenses share the same optical axis perpendicular to the window. In order to accommodate an optical axis perpendicular to the window, a prism is used, which for this concept is made of alumina, although it can be made out of silicon. This prism could in principle be combined with the first lens in the optical chain, which would allow for a more compact design with a higher throughput. The choice of orthogonality of the camera optical axis can be relaxed in future iterations of this design in order to relieve the optical power of the alumina prism. The implementation of these options are left for future studies.

\begin{figure}[bth] 
    \begin{minipage}{.65\textwidth}
        \centering
        \includegraphics[angle=-90, width=\textwidth, trim={2.7cm 0.2cm 0.3cm 2.85cm},clip]{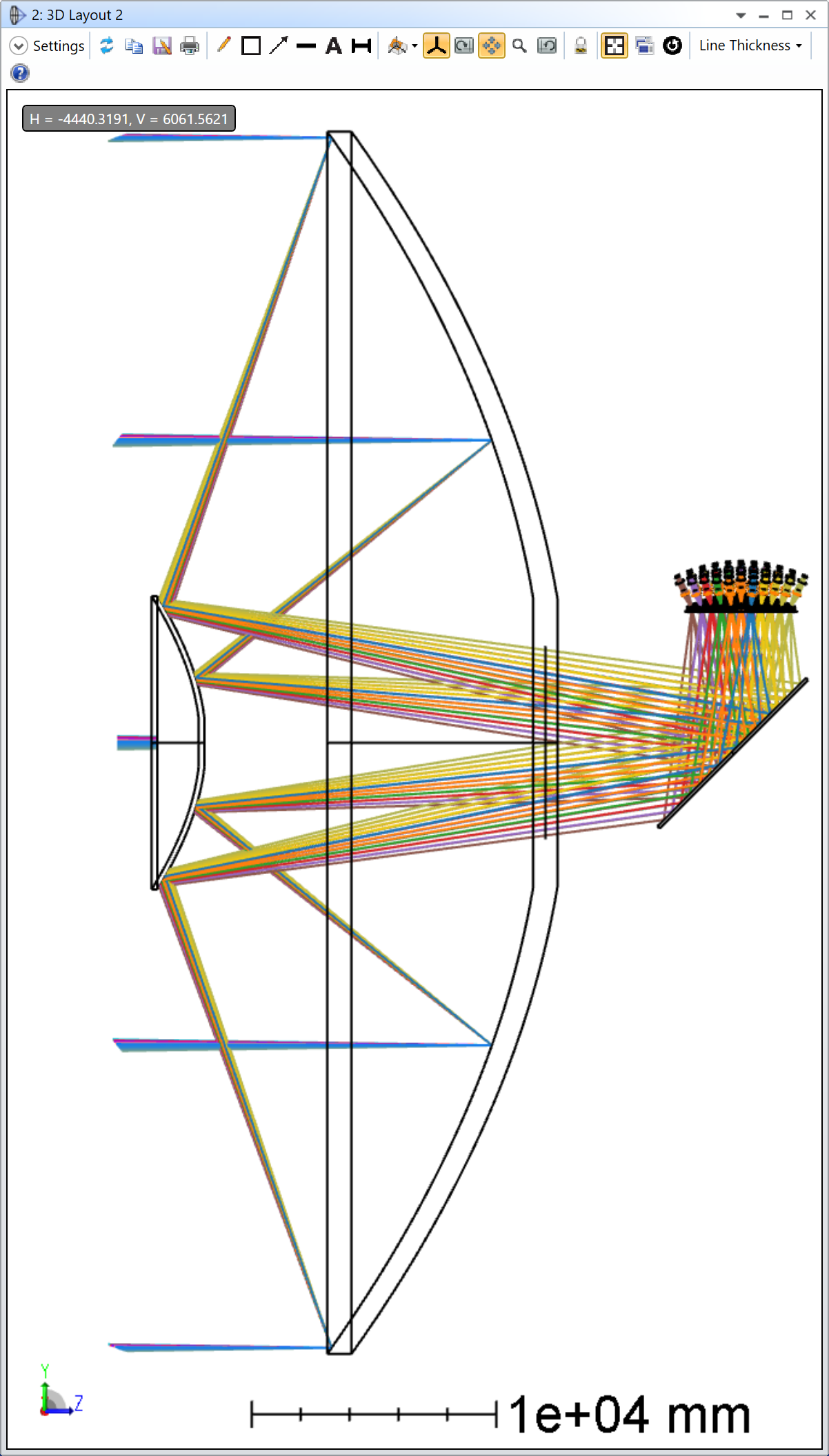}
        \label{fig:prob1_6_2}
    \end{minipage}%
    \begin{minipage}{0.35\textwidth}
        \centering
        \includegraphics[width=\textwidth, trim={2cm 0.2cm 3cm 2cm},clip]{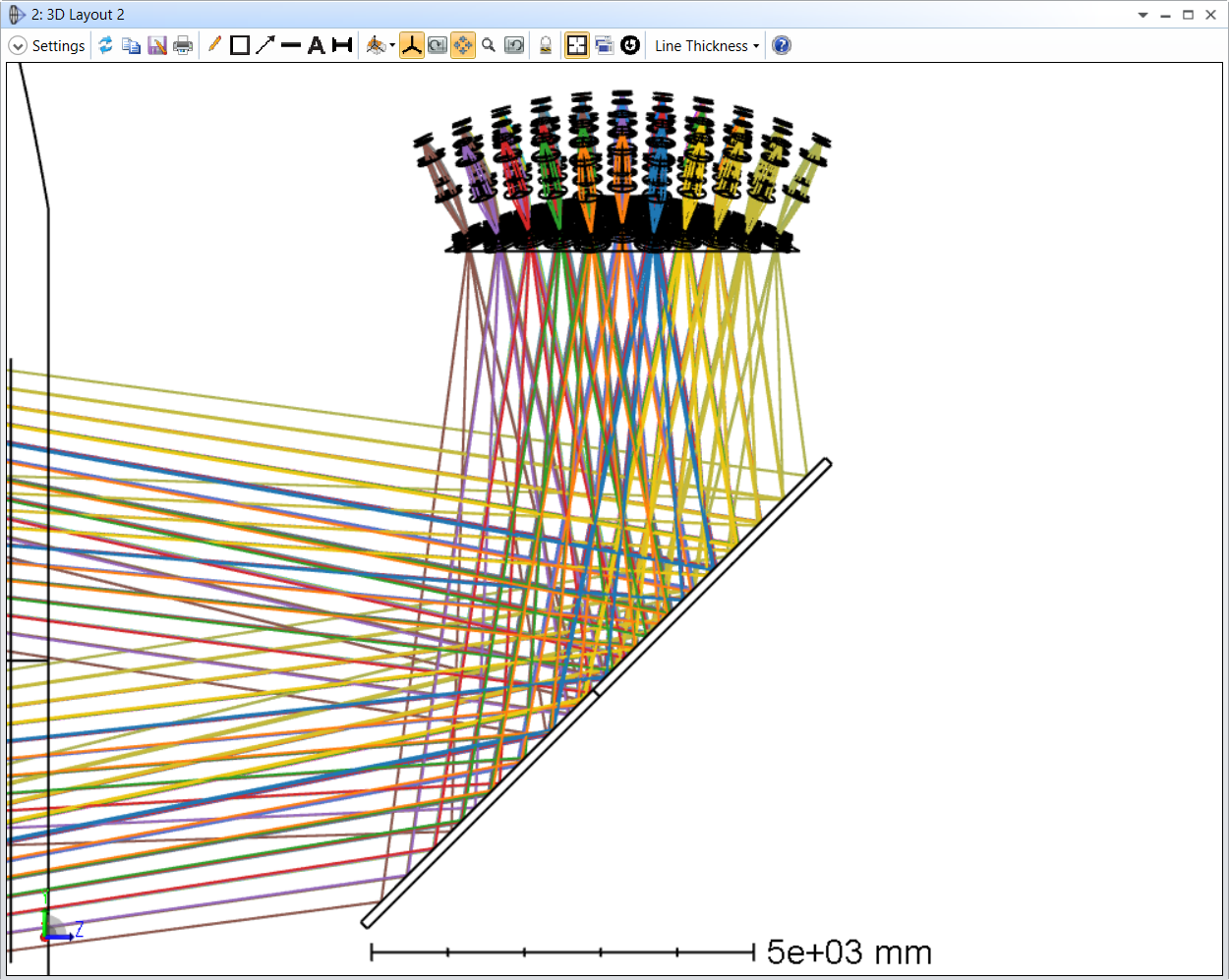}
        \label{fig:prob1_6_1}
    \end{minipage}
    \caption{Baseline camera concept for AtLAST. This camera system is composed of 85 cameras, with a diameter of 40 cm each. The windows follow the shape of the focal surface to keep the camera on axis. A prism is used to redirect the beam towards the axis of the camera. This camera concept is composed of three lenses and a Lyot stop. The second lens in the system can follow a biconic shape, which corrects for astigmatism. Optical path from primary mirror is shown on the left panel, right panel shows a zoom-in view of the 85 cameras.}
    \label{fig:atlastmulticam}
\end{figure}

Figure~\ref{fig:atlastmulticam} shows the 85 cameras in the camera concept, which makes use of a biconic lens to correct for the dominant optical aberration in the two-mirror telescope. The optimal location to correct for astigmatism in this camera concept is the Lyot stop, however, making the second lens a biconic surface makes the system simpler and attenuates the risk of possible manufacturing defects, which have the potential to contaminate the beam shape. The biconic surface of the second lens has a shape according to \begin{equation}z(x,y) = \frac{x^2/R_x + y^2/R_y}{1+\sqrt{1-(1+k_x)(x/R_x)^2 - (1+k_y)(y/R_y)^2}} ,
\end{equation}
 where $R_x$ and $R_y$ are the radii of curvature, and $k_x$ and $k_y$ are the conic constants along the \emph{x, y} axes. The local system of coordinates for this lens is rotated around the optical axis to find the best orientation to cancel the astigmatism. The first and third lens are made of a rotationally symmetric standard surface.

\begin{figure}[tbh]
    \centering
    \includegraphics[width=0.495\textwidth]{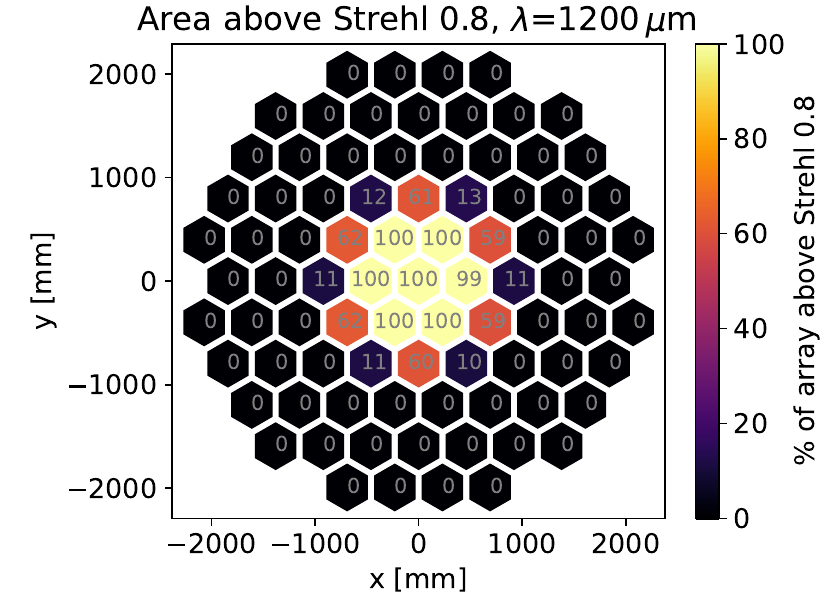}
    \includegraphics[width=0.495\textwidth]{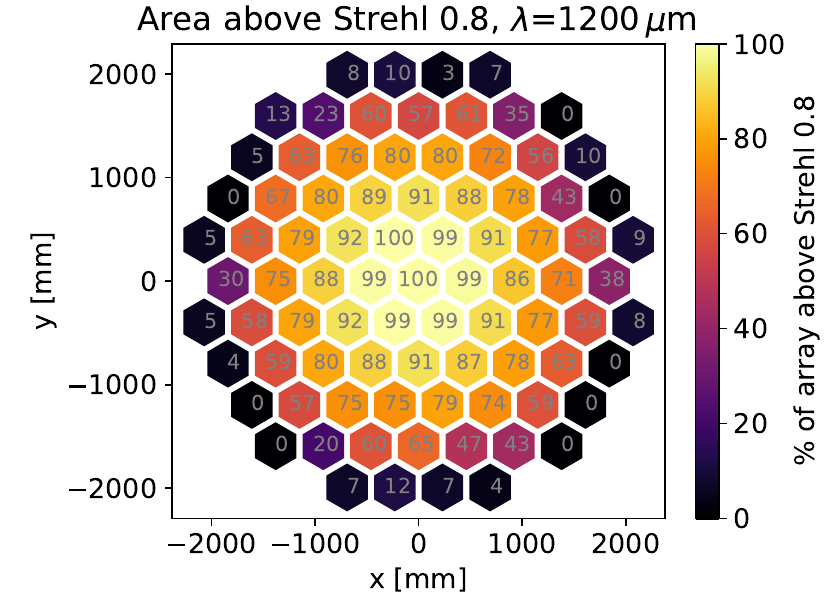}

    \includegraphics[width=0.495\textwidth]{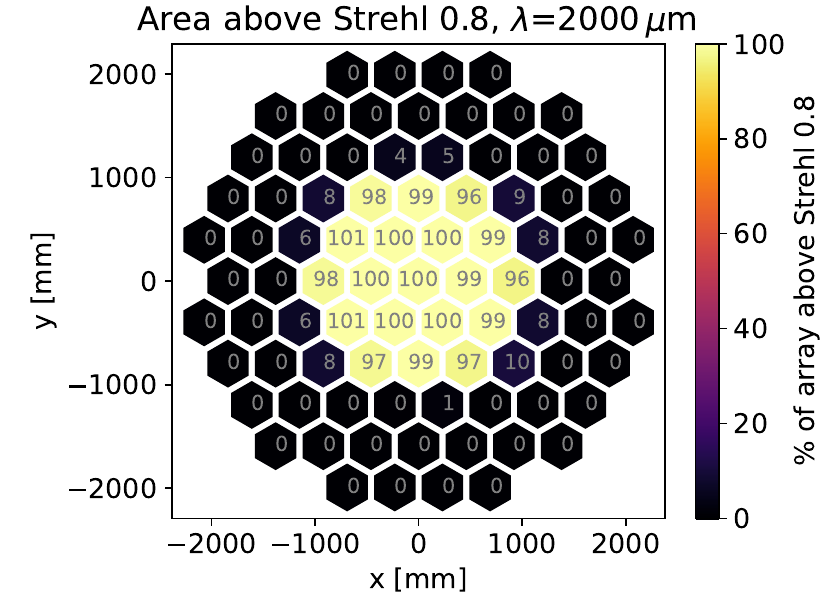}
    \includegraphics[width=0.495\textwidth]{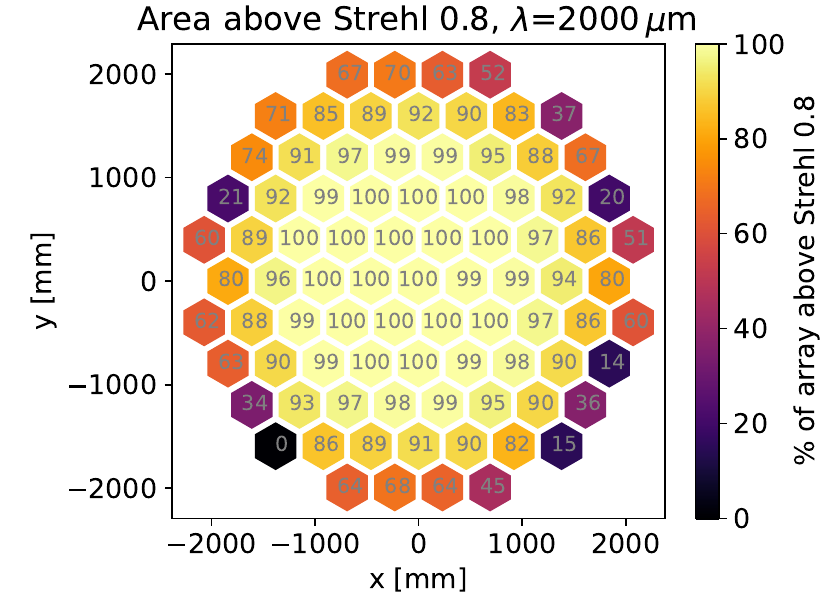}
    
    \includegraphics[width=0.495\textwidth]{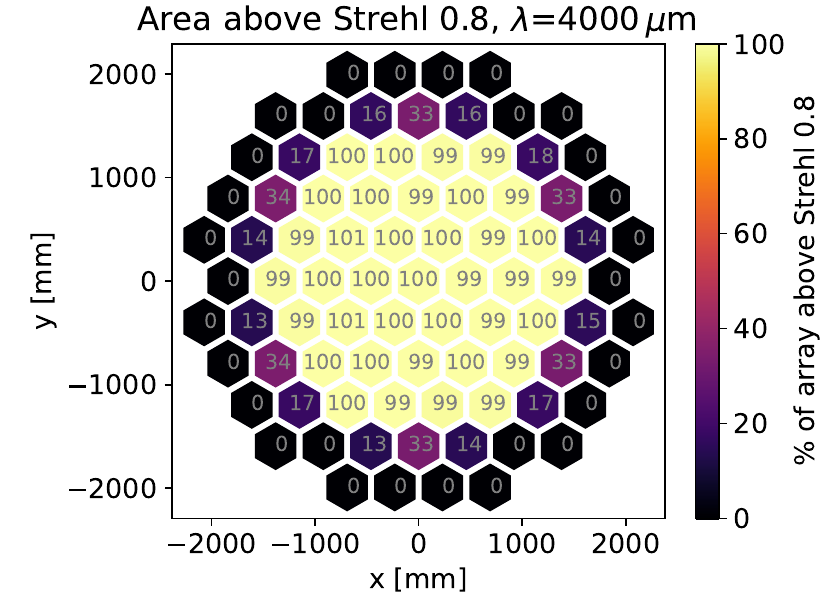}
    \includegraphics[width=0.495\textwidth]{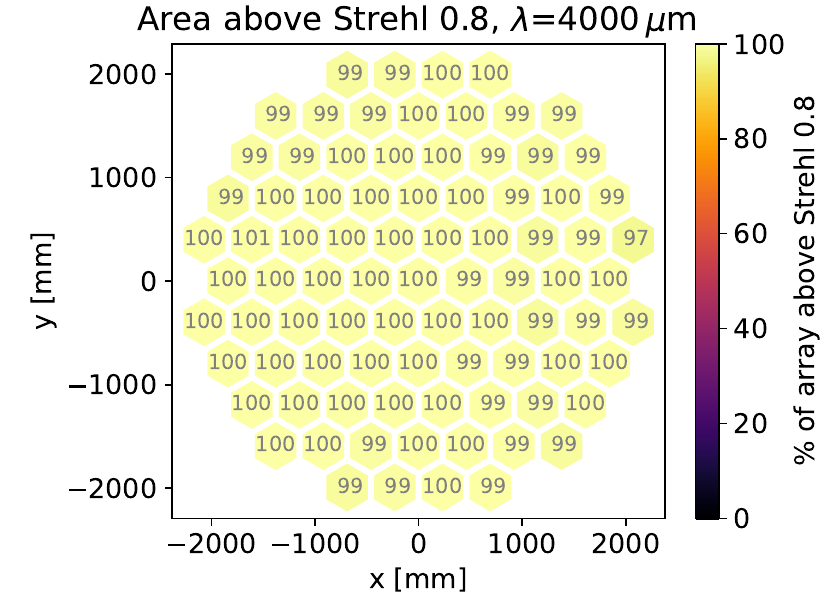}
    \caption{Image quality for an 85-camera concept (Fig.~\ref{fig:atlastmulticam}) populating the 2-degree field of view of AtLAST. Lenses are $40\, \rm cm$ in diameter, which allows to populate the 2-degree focal plane with minimal gaps. Figure shows 85 hexagons which show the percentage of the detector focal plane with a Strehl ratio higher than 0.8. Left column shows this metric for a radially symmetric second lens in the camera system, while the right column shows the same metric for a biconic second lens. Diffraction-limited performance can be achieved at 4 millimeters and at shorter wavelengths the corrector lens allows a significant improvement of the usable field of view.}
    \label{fig:opticalqual40cmcams}
\end{figure}

The array of 85 cameras presented here with a biconic lens is compared to a similar system with a rotationally symmetric second lens in Figure~\ref{fig:opticalqual40cmcams} for three wavelengths of interest (1.2\,mm, 2.0\,mm and 4\,mm). The column of the left shows the percentage of the focal plane with a Strehl ratio higher than 0.8 for a radially symmetric second lens, while the column of the right shows the same metric for a three-lens system with a biconic second lens. The top row shows that a rotationally symmetric camera would yield seven diffraction-limited cameras at $1.2\, \rm mm$, however by including a biconic lens, a second ring of 7 additional cameras can be used with diffraction-limited performance, in addition, the optical quality degrades slowly towards the edges, which is an indication that the optical quality can be improved further with detailed optimization, or that cameras at these locations can still be used with a somewhat degraded performance. The second row shows that at $2\,\rm mm$ of wavelength, the system becomes diffraction-limited in more than 95\% of the array in 38 cameras, except for the two outermost rings. Here it can be seen that the second to last ring of cameras still has a coverage with more than $80\,\%$ of the focal plane showing diffraction-limited performance. The last row shows that at a wavelength $\geq 4\, \rm mm$, the full 2 degree field of AtLAST can be used with diffraction-limited performance.

\begin{figure}
    \centering
    \includegraphics[width=0.495\textwidth]{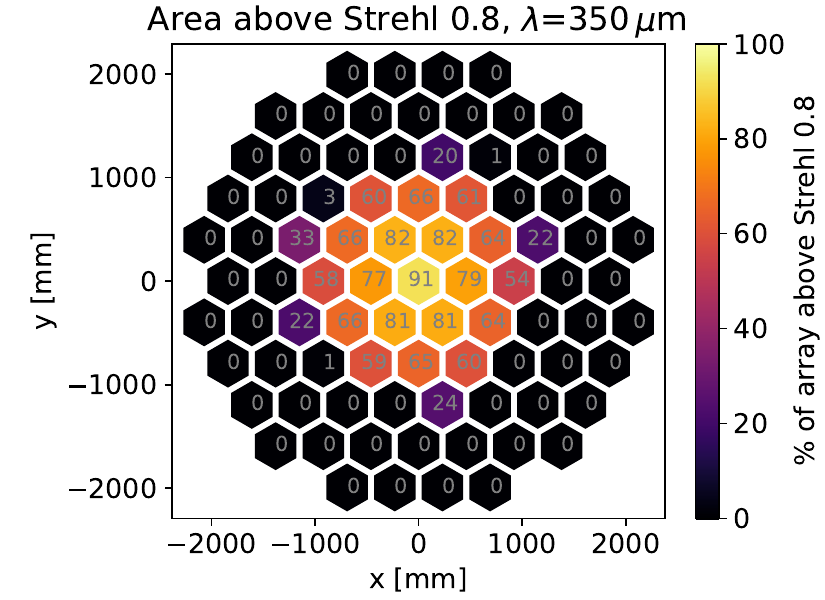}
    \includegraphics[width=0.495\textwidth]{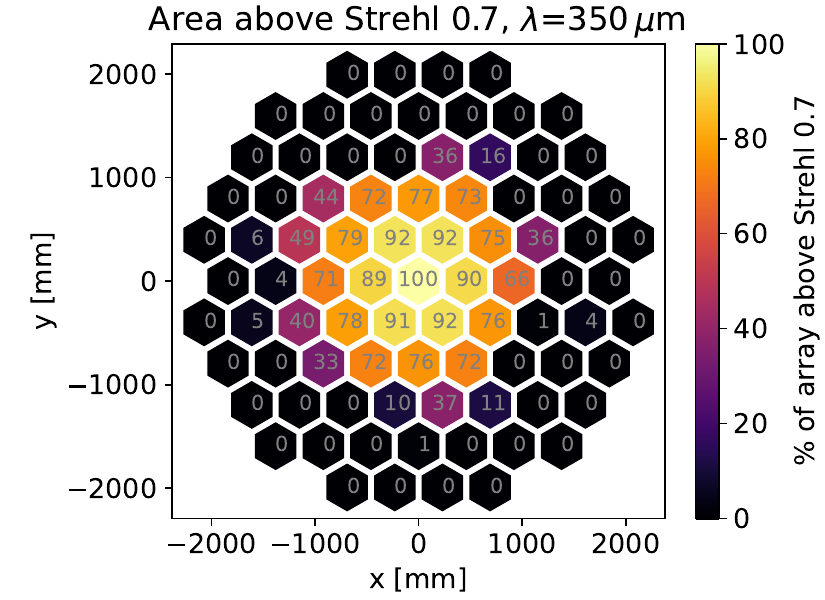}
    \includegraphics[width=0.495\textwidth]{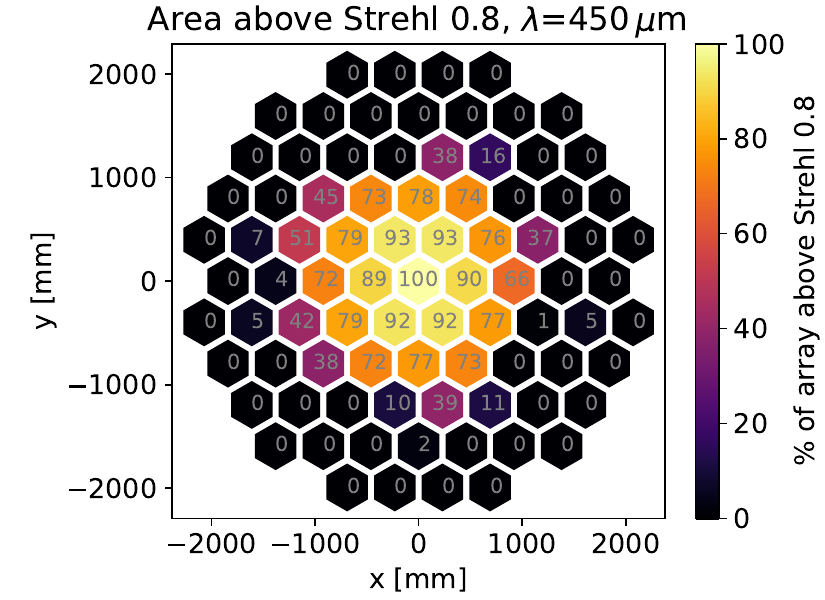}
    \includegraphics[width=0.495\textwidth]{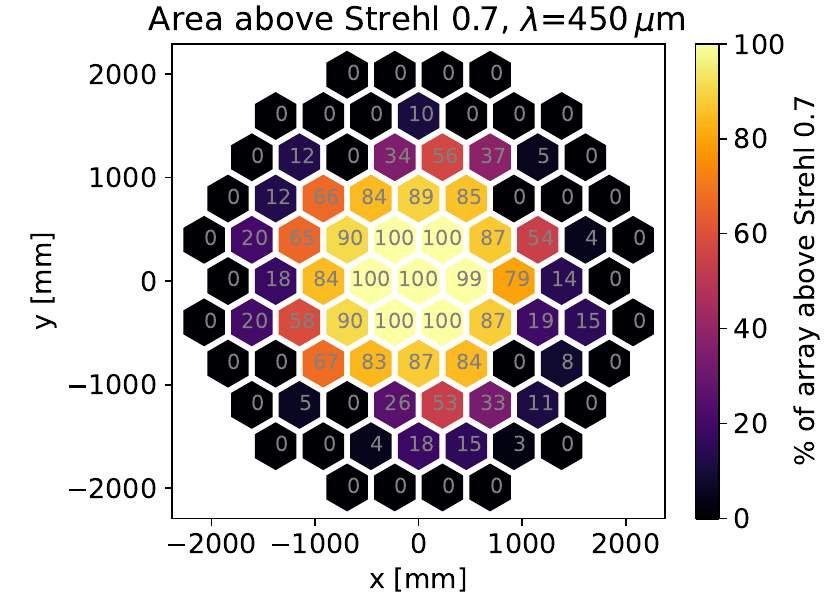}
    \includegraphics[width=0.495\textwidth]{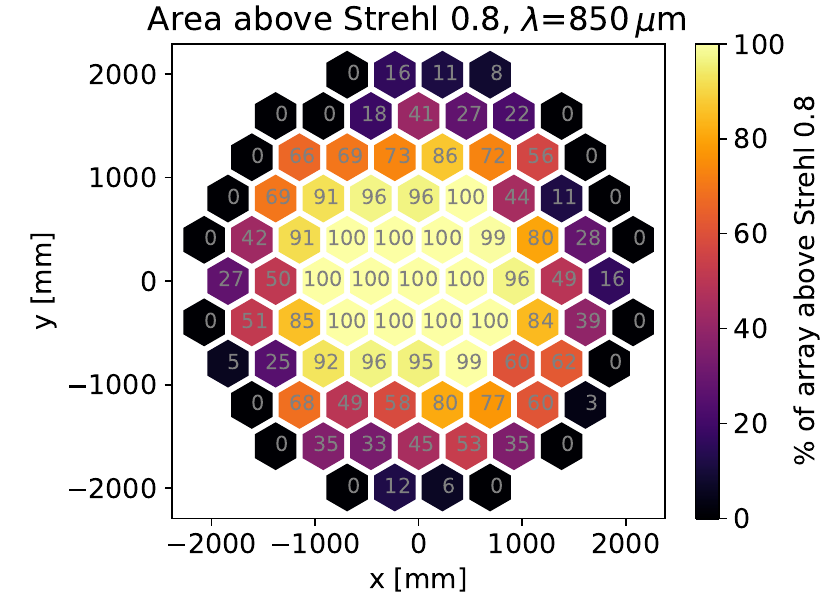}
    \includegraphics[width=0.495\textwidth]{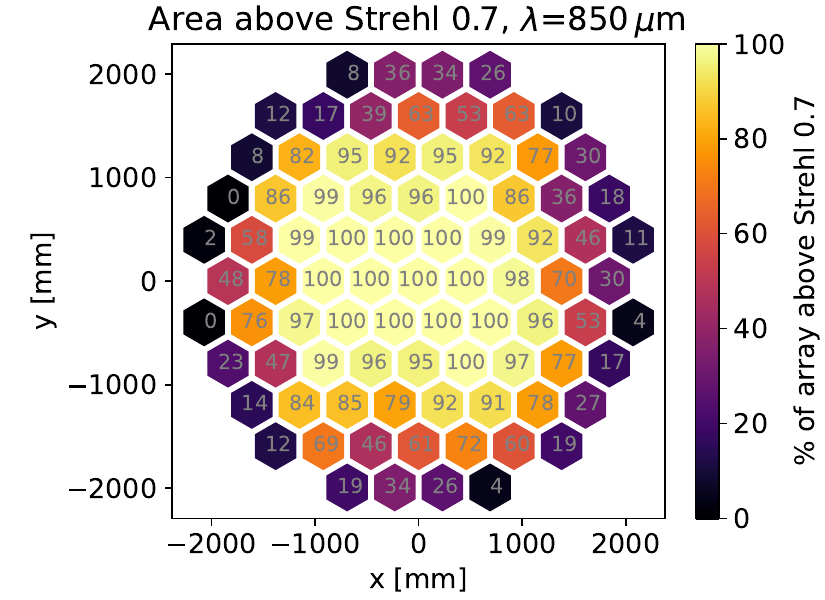}
    \caption{Optical quality for 85 cameras made using $20\, \rm cm$ diameter lenses. Figure shows this system at wavelengths of $350\, \rm \mu m$, $450\, \rm \mu m$ and $850\, \rm \mu m$. Figure~\ref{fig:mm_20cm} shows performance for millimeter wavelengths. Cameras were optimized individually. Asymmetry in this diagram is the result of numerical noise, and can be improved by forcing the system to be radially symmetric.}
    \label{fig:submm_20cm}
\end{figure}

\begin{figure}
    \centering
    \includegraphics[width=0.495\textwidth]{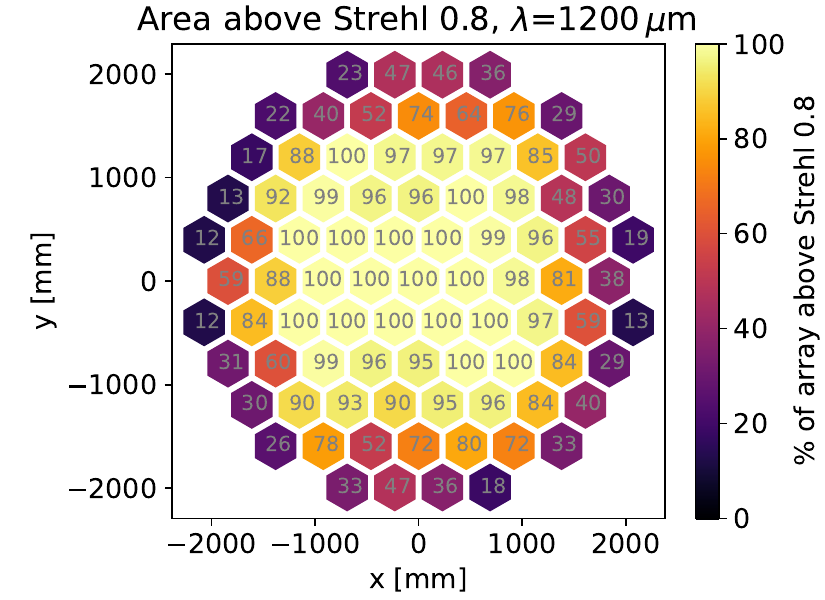}
    \includegraphics[width=0.495\textwidth]{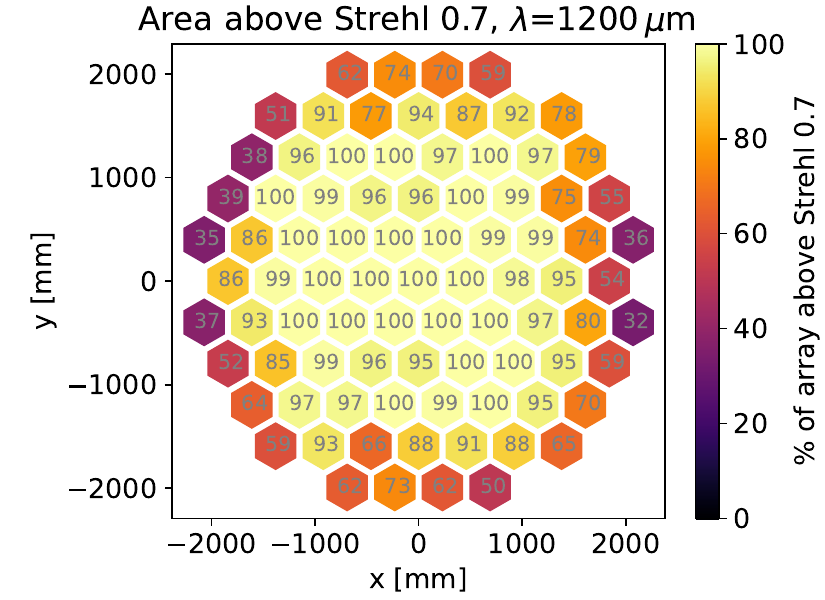}
    \includegraphics[width=0.495\textwidth]{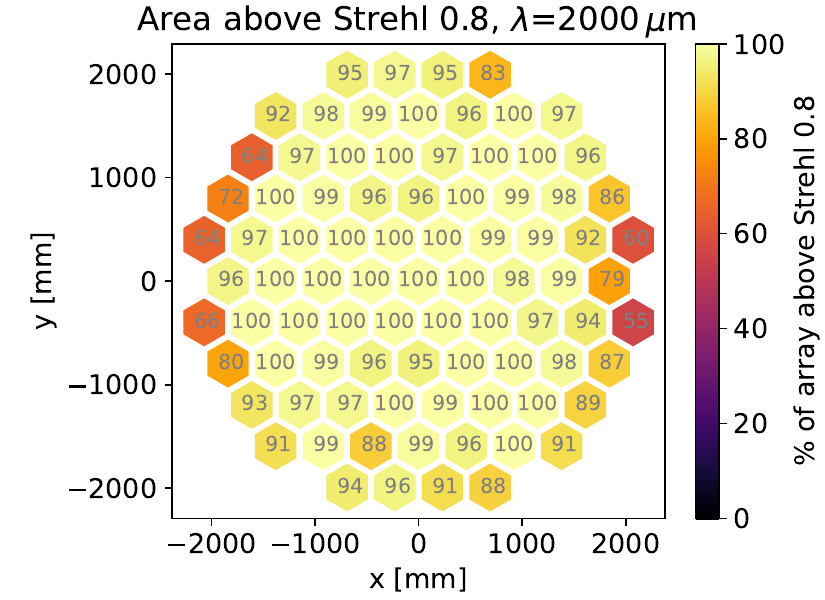}
    \includegraphics[width=0.495\textwidth]{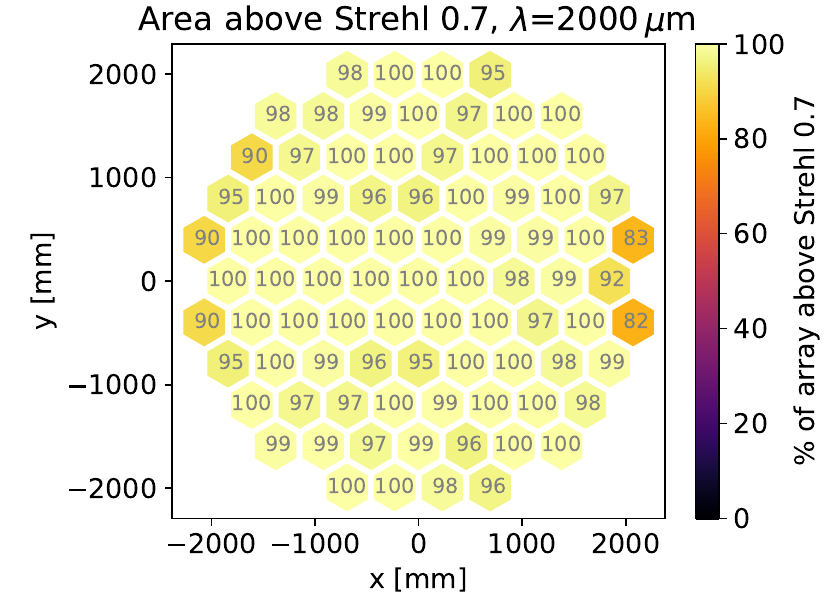}
    \caption{Optical quality for 85 cameras made using $20\, \rm cm$ diameter lenses. Figure shows this system at wavelengths of $1.2\, \rm  mm$ and $2.0\, mm$. Cameras were optimized individually. Asymmetry in this diagram is the result of numerical noise, and can be improved by forcing the system to be radially symmetric.}
    \label{fig:mm_20cm}
\end{figure}

An additional system with smaller 20~cm diameter lenses is shown in Figures \ref{fig:submm_20cm} and \ref{fig:mm_20cm}. The columns show the percentage of the camera focal plane with Strehl ratios higher than 0.8 (left) and 0.7 (right). At $350\, \rm \mu m$ the central seven hexagons (0.4~degrees) show diffraction-limited performance in more than $77\%$ of the array, and the central 19 cameras (0.8~degrees) can be populated if the Strehl ratios are relaxed to 0.7.  At $450 \, \rm \mu m$ this coverage remains unchanged, with better Strehl ratios in the 19 cameras. At $850 \, \rm \mu m$, 1.2 degrees of field of view can be used, which increases to 1.83 degrees (4.2 meters) for the full field of view if the Strehl ratios are relaxed to 0.7. At still longer wavelengths ($\lambda > 1.2$~mm), instruments can more easily fill the entire 2-degree field of view (see Figure~\ref{fig:mm_20cm}) without relaxing the Strehl ratio requirements.

\section{Conclusions}\label{sec:conclusions}
AtLAST is proposed to be the premier facility for single-dish millimeter and sub-millimeter astronomy. We presented the optical design of AtLAST, which consists of a fifty-meter primary mirror and a twelve-meter secondary mirror with a fold mirror, which allows users to select among six different focal surfaces. An optical demonstration for an array of three-lens biconic cameras is presented, which can be used as a starting point for the future development of instruments to deliver the highest throughput large dish telescope spanning the millimeter and sub-millimeter spectrum.

\acknowledgments 

This project has received funding from the European Union’s Horizon 2020 research and innovation program under grant agreement No.\ 951815 (AtLAST).\footnote{See
\href{https://cordis.europa.eu/project/id/951815}{https://cordis.europa.eu/project/id/951815}.} The design presented here evolved from pioneering work by R. Hills on various alternative optics designs \cite{AtLAST_memo_1}. PAG acknowledges support from KICP at U. Chicago.

\bibliography{report} 
\bibliographystyle{spiebib} 

\end{document}